\documentclass[epj]{svjour}
\usepackage{graphicx}

\begin{document}
   
 
\title{Mean-field model for Josephson oscillation in 
a Bose-Einstein condensate on an 
one-dimensional
optical trap}
\author{Sadhan K. Adhikari\thanks{e-mail: adhikari@ift.unesp.br}}
\institute{Instituto de F\'{\i}sica Te\'orica, Universidade Estadual
Paulista, 01.405-900 S\~ao Paulo, S\~ao Paulo, Brazil}



\date{Received: 29 January 2003 / Revised: 20 March 2003 / Accepted: 20
March 2003}       

\abstract{ Using the axially-symmetric time-dependent Gross-Pitaevskii
equation we study
the phase
coherence in a repulsive Bose-Einstein condensate (BEC)  trapped by a harmonic and an
one-dimensional optical lattice potential to describe the experiment by
Cataliotti {\it et al.} on atomic Josephson oscillation [Science {\bf
293}, 843 (2001)].  The phase
coherence is maintained after the BEC is set into oscillation by a small
displacement of the magnetic trap along the optical lattice.  The phase
coherence in the presence of oscillating neutral current across an array
of Josephson junctions manifests in an interference pattern formed upon
free expansion of the BEC. The numerical response of the system to a large
displacement of the magnetic trap is a classical transition from a
coherent superfluid to an insulator regime and a subsequent destruction of
the interference pattern in agreement with the more 
recent experiment by
Cataliotti {\it et al.} [e-print cond-mat/0207139]. }


\authorrunning{S. K. Adhikari}
\titlerunning{Josephson oscillation in a Bose-Einstein condensate on an
optical trap}
\maketitle

\section{Introduction}
 
The observation of an oscillating Josephson current across the boundaries
of a one-dimensional
array of potential wells, usually generated by a standing-wave laser
field and  
commonly known as an  optical lattice potential, in a trapped
cigar-shaped Bose-Einstein condensate (BEC) by Cataliotti {\it et al.}
\cite{cata} is a clear  manifestation of macroscopic quantum
phase coherence.  So far the 
Josephson effect has been confirmed in superconductors with charged
electrons and in liquid helium
\cite{3}.

The recent experimental observation of BEC in trapped alkali-metal atoms
\cite{exp} has offered new possibility of the confirmation of Josephson
effect in neutral quantum fluids with an array of quasi
one-dimensional Josephson junctions not
realizable in superconductors. The experimental loading of a cigar-shaped
BEC in  both one- \cite{1,2,ari} and three-dimensional \cite{greiner}
optical
lattice potentials
has allowed the study of quantum phase
effects on a macroscopic scale such as interference of matter
waves \cite{kett1}.  There have been several theoretical studies on
different
aspects of a BEC in a one-  \cite{th} as well as three-dimensional
\cite{adhi}
optical lattice
potentials.
The phase coherence between
different sites of a trapped BEC on an optical lattice has been
established in  recent experiments
\cite{cata,1,2,greiner} through the formation of
distinct interference pattern when the traps are removed.

Cataliotti {\it et al.} \cite{cata,cata3b} have provided a quantitative
measurement
of the formation and evolution of interference pattern upon free expansion
of a cigar-shaped trapped BEC of repulsive Rb atoms on an optical lattice
and harmonic potentials 
after the removal of the combined traps. 
 The phase coherence in
a BEC trapped in a standing-wave
optical-lattice is responsible for the formation
of a distinct interference pattern upon free expansion as observed in
several recent experiments \cite{cata,1,2,greiner,cata3b}. Cataliotti
{\it et al.}
\cite{cata} also
continued their
investigation to a BEC oscillating on the optical lattice via quantum
tunneling and found that the phase coherence between different sites is
maintained during oscillation initiated by a sudden shift of the
magnetic trap  along the optical axis. 

The
phase-coherent BEC on the optical lattice is a quantum superfluid
\cite{greiner}
and the atoms in
it move freely from one optical site to another by quantum tunneling.
However, 
the classical movement is prohibited through the high optical potential
traps. It has been demonstrated in a recent experiment by Greiner {\it et
al.} \cite{greiner}  
that, as the optical potential traps are made much too higher, the
quantum tunneling of atoms from one optical site to another is stopped
resulting in a loss of superfluidity and phase coherence in the BEC.
Consequently, no interference pattern is formed upon free expansion of
such a BEC which is termed a Mott insulator state. This phenomenon
represents a quantum phase transition (with energy nonconservation in the
tunneling process) and cannot be accounted for in a classical mean-field
model
based
on the Gross-Pitaevskii (GP) equation \cite{8}.

Following a suggestion by Smerzi {\it et al.} \cite{sm}, more recently
Cataliotti
{\it et al.} \cite{cata2} have demonstrated in a novel experiment the loss
of phase
coherence and superfluidity in a BEC trapped in a optical-lattice and 
harmonic potentials when the center of the harmonic potential is suddenly
displaced along the optical lattice through a distance larger
than a critical value. 
Then  a modulational instability takes place in the
BEC and it  cannot reorganize itself quickly enough and the
phase coherence and superfluidity of the BEC are destroyed. The
resulting motion of
the condensate is not oscillatory in nature. 
The loss of phase coherence  is
manifested in the destruction of the interference pattern upon free
expansion. However, for displacements smaller than the critical distance
the BEC can reorganize itself and the phase coherence and superfluidity
are maintained \cite{cata,cata2}. Recently, a new classical mechanism for
the loss of superfluidity of a BEC arising from a nonlinear modulation of
the scattering  length has been suggested \cite{loss}.
Distinct from the quantum phase transition observed by
Greiner
{\it et al.} \cite{greiner}, these modulational instabilities responsible
for
the destruction
of phase coherence are  classical in nature and  can be described
\cite{sm,cata2} by
the mean-field model. Hence in the present paper we present a mean-field
description of the experiments by Cataliotti {\it et al.}
\cite{cata,cata2}
to see how well
and how far it can describe the observed phenomena. Specifically, we
consider the numerical solution of the axially-symmetric GP equation
\cite{8}
 in an 
optical and a harmonic trap.

Cataliotti {\it et al.} \cite{cata,cata2} provided a  
theoretical account of
their study using   the tight-binding approximation for the full
wave function 
in the presence of the
periodic optical potential wells. Also, there has been a preliminary
attempt to explain some features of this experiment using 
one-dimensional mean-field models \cite{cata3b,cata3}. In
reference \cite{cata3} a
dynamical solution of one-dimensional GP equation was used; whereas in
reference 
\cite{cata3b}  an one-dimensional model of interference was developed
using superposition of analytical matter waves, which is reasonable in the 
absence of nonlinear atomic interaction.   
Although, the tight-binding approximation and these 
one-dimensional models
could be reasonable for the study of some aspects of the experiment of
Cataliotti {\it et al.} \cite{cata,cata2},
here we  compare the results  with  the complete
solution of the three-dimensional mean-field Hamiltonian via the nonlinear
GP equation \cite{8}.

In section 2  we present the mean-field model based on the
axially-symmetric time-dependent nonlinear 
GP equation. In section 3 we present the numerical results and finally, in
section 4 we present the conclusions.

\section{Mean-field Model}

The time-dependent BEC wave
function $\Psi({\bf r};t)$ at position ${\bf r} $ and time $t $
is described by the following  mean-field nonlinear GP equation
\cite{8}
\begin{eqnarray}\label{a} \left[- i\hbar\frac{\partial
}{\partial t}
-\frac{\hbar^2\nabla^2   }{2m}
+ V({\bf r})
+ gN|\Psi({\bf
r};t)|^2
 \right]\Psi({\bf r};t)=0,
\end{eqnarray}
where $m$
is
the mass and  $N$ the number of atoms in the
condensate,
 $g=4\pi \hbar^2 a/m $ the strength of interatomic interaction, with
$a$ the atomic scattering length.  In the presence of the combined
axially-symmetric and optical lattice traps 
     $  V({\bf
r}) =\frac{1}{2}m \omega ^2(\rho ^2+\nu^2 y^2) +V_{\mbox{opt}}$ where
 $\omega$ is the angular frequency of the harmonic trap 
in the radial direction $\rho$,
$\nu \omega$ that in  the
axial direction $y$, with $\nu$ the aspect ratio, 
and $V_{\mbox{opt}}$ is
the optical lattice potential  introduced later.  
The normalization condition  is
$ \int d{\bf r} |\Psi({\bf r};t)|^2 = 1. $

In the axially-symmetric configuration, the wave function
can be written as 
$\Psi({\bf r}, t)= \psi(\rho,y,t)$, where $0\le  \rho < \infty$ is the
radial
variable and $-\infty <y<\infty $ is the axial variable.
Now  transforming to
dimensionless variables $\hat \rho =\sqrt 2 \rho/l$,  
$\hat y=\sqrt 2 y/l$,   $\tau
=t
\omega, $
$l\equiv \sqrt {\hbar/(m\omega)}$,
and
${ \varphi(\hat \rho,\hat y;\tau)} \equiv   \hat \rho \sqrt{{l^3}/{\sqrt
8}}\psi(\rho,y;t),$   equation (\ref{a}) becomes \cite{9}
\begin{eqnarray}\label{d1}
&\biggr[&-i\frac{\partial
}{\partial \tau} -\frac{\partial^2}{\partial
\hat \rho ^2}+\frac{1}{\hat \rho }\frac{\partial}{\partial \hat \rho}
-\frac{\partial^2}{\partial
\hat y^2}
+\frac{1}{4}\left(\hat \rho ^2+\nu^2 \hat y^2\right) \nonumber \\
&+&\frac{V_{\mbox{opt}}}{\hbar \omega} -{1\over \hat \rho ^2}  +
   8\sqrt 2 \pi n\left|\frac {\varphi({\hat \rho ,\hat y};\tau)}{\hat
\rho}\right|^2
 \biggr]\varphi({ \hat \rho,\hat y};\tau)=0, 
\end{eqnarray}
where
$ n =   N a /l$. In terms of the 
one-dimensional probability 
 $P(y,t) \equiv 2\pi\- \- \int_0 ^\infty 
d\hat \rho |\varphi(\hat \rho,\hat y,\tau)|^2/\hat \rho $, the
normalization of the
wave 
function 
is given by $\int_{-\infty}^\infty d\hat y P(y,t) = 1.$  The probability 
$P(y,t)$ is  useful in the study of the present problem under the
action of the optical lattice potential, specially in the
investigation of the formation and
evolution of the interference pattern after the removal of the
trapping potentials.

In the  experiment of Cataliotti {\it et al.} \cite{cata}
with repulsive $^{87}$Rb atoms in the hyperfine state $F=1,
m_F=-1$, the axial and radial trap frequencies were $\nu \omega =
2\pi \times 9 $ Hz and $ \omega =
2\pi \times 92$ Hz, respectively, with $\nu = 9/92$. The
optical
potential created with the standing-wave laser field of wavelength 
$\lambda=795$ nm is given by $V_{\mbox{opt}}=V_0E_R\cos^2 (k_Lz)$,
with $E_R=\hbar^2k_L^2/(2m)$, $k_L=2\pi/\lambda$, and $V_0$ $ (<12)$ the 
strength. For the mass $m=1.441 \times 10^{-25}$ kg of $^{87}$Rb the
harmonic
oscillator length $l=\sqrt {\hbar/(m\omega)} = 1.126$ $\mu$m and the
present 
dimensionless length unit  corresponds to $l/\sqrt 2 =0.796$ $\mu$m. The
present
dimensionless time unit corresponds to $\omega ^{-1} =
1/(2\pi\times 92)$ s $=1.73$ ms. Although we perform the calculation in
dimensionless units using equation (\ref{d1}), we present the results in
actual physcial units using these conversion factors consistent with the
experiment by Cataliotti {\it et al.} \cite{cata}.
In terms of the dimensionless laser wave
length $\lambda _0= \sqrt2\lambda/l \simeq 1$, the dimensionless 
standing-wave energy parameter $E_R/(\hbar \omega)= 4\pi^2/\lambda _0^2$.
Hence in 
dimensionless unit $V_{\mbox{opt}}$ of 
equation   (\ref{d1}) is
\begin{equation}\label{pot}
\frac{ V_{\mbox{opt}}}{\hbar \omega}=V_0\frac{4\pi^2}{\lambda_0^2} 
\left[
\cos^2 \left(
\frac{2\pi}{\lambda_0}\hat y
\right)
 \right].
\end{equation}

We solve  equation  (\ref{d1}) numerically  using a   
split-step time-iteration
method
with  the Crank-Nicholson discretization scheme described recently
\cite{11}.  
The time iteration is started with the known harmonic oscillator solution
of  equation  (\ref{d1}) with
 $n=0$: $\varphi(\hat \rho, \hat y) = [\nu
/(8\pi^3)  ]^{1/4}$
$\hat \rho$ $e^{-(\hat \rho^2+\nu \hat y  ^2)/4}$ with chemical potential
$\bar
\mu=(1+\nu /2)$
\cite{9}. For a typical cigar-shaped condensate with $\nu \simeq 0.1$
\cite{cata} $\bar \mu \simeq 1$ is much smaller than the typical depth of
the 
optical potential wells $E_R/(\hbar \omega) = 4\pi^2 /\lambda _0^2 \simeq
40$  so that $\bar \mu
<<
E_R/(\hbar \omega)$ and the  passage of condensate atoms from one well to
other can only
proceed through quantum tunneling. 
The
nonlinearity  as well as the optical lattice potential parameter $V_0$ 
are  slowly increased by equal amounts in $10000n$ steps of 
time iteration until the desired value of nonlinearity and optical lattice
potentials are  attained. Then, without changing any
parameter, the solution so obtained is iterated 50 000 times so that a
stable
solution  is obtained 
independent of the initial input
and time and space steps. 
The
solution then corresponds to the bound BEC under the joint action of
the harmonic and optical traps.

\begin{figure}
 
\begin{center}

\includegraphics[width=1.\linewidth]{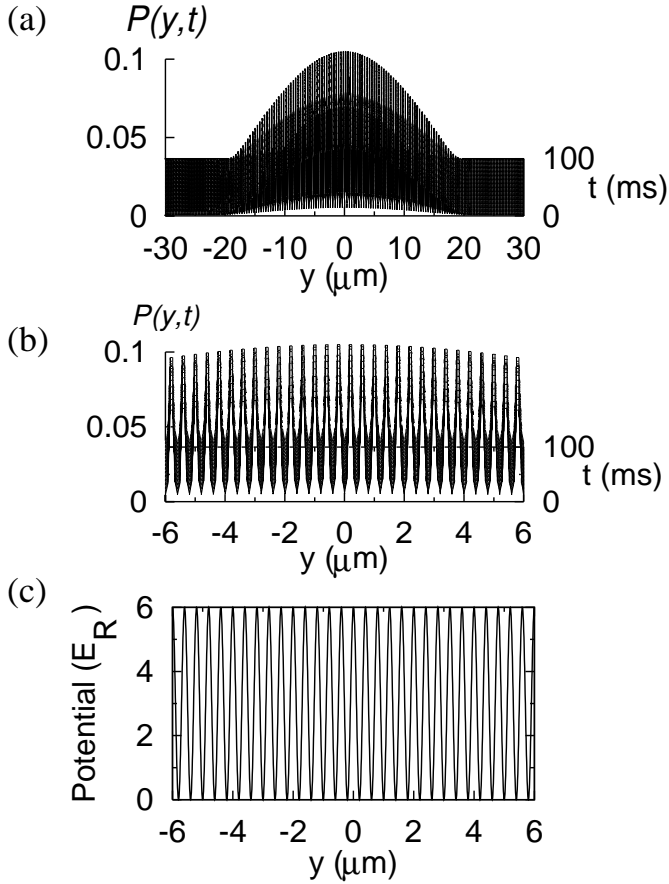}
\end{center}
 
\caption{$P(y,t)$
vs. $y$ and $t$
for the ground-state BEC with $n=10$ and $V_0=6$  for $0<t<100$ ms and
(a) $-30$ $\mu\mbox{m}<y<30$ $\mu\mbox{m}$ and
(b) $-6$ $\mu\mbox{m}<y<6$ $\mu\mbox{m}$. The optical potential for
$-6$ $\mu\mbox{m}<y<6$ $\mu\mbox{m}$
is shown in (c). The harmonic oscillator potential is negligible on this
scale. 
} \end{figure}

\begin{figure}
 
\begin{center}
\includegraphics[width=1.\linewidth]{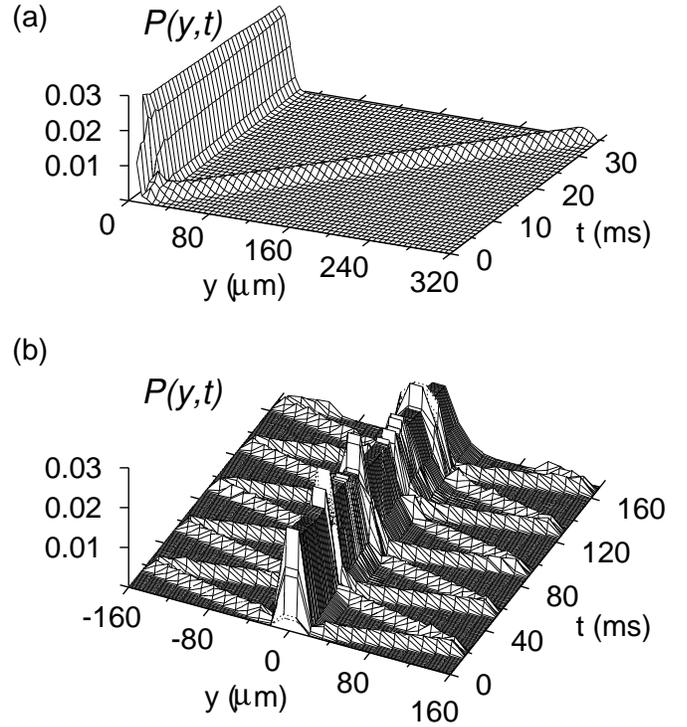}
\end{center}
 
\caption{$P(y,t)$  
vs. $y$ and $t$ for the BEC of figure 1 after the removal of 
combined traps at $t=0$ for a lattice defined by $\rho \le 25$
$\mu\mbox{m}$
and
(a) $-320$  $\mu\mbox{m}<y<320$ $\mu\mbox{m}$ and (b)  $-160$
$\mu\mbox{m}<y<160$  $\mu\mbox{m}$.}
\end{figure}

\section{Numerical Results} 

First we consider the  BEC formed on the
optical
lattice  for a specific nonlinearity. In the present study
we take  nonlinearity $n=10$ and  optical lattice strength 
$V_0=6$ except in figure 5 where we use a variable $V_0$. We consider the
ground-state wave
function in the combined harmonic  and optical lattice
potentials.  The one-dimensional pattern in
the axial $y$ direction is most easily illustrated from a consideration of
the
probability $P(y,t)$ at different
times. In figure 1 (a) we plot the frontal view of  $P(y,t)$ for 
 $0<t<100$ ms and $-30 \mu\mbox{m} <y<30 \mu\mbox{m}$. In this interval of
$y$, there are 150
wells of the optical potential and as many maxima and minima in  $P(y,t)$,
which cannot be visualized clearly  in figure 1 (a). In the actual
experiment 200 wells were typically populated, which corresponds to a
larger condensate than considered in this numerical simulation.  For the
limitation in 
computer processing time  we had to stick to a  
smaller
condensate. 
In figure 
1
(b) we
show a close-up of figure  1 (a) for $-6$ $\mu\mbox{m}<y<6$ $\mu\mbox{m}$
containing 30 wells. The
corresponding optical
potential   is shown in figure  1 (c), which clearly shows the 30 wells. 
From
figure  1 (b) one can count  30 maxima and 30 minima in 
probability  $P(y,t)$.

As the present calculation is performed with the full wave function
without approximation, phase coherence among different wells of the 
optical lattice   is automatically guaranteed. As a result when
the
condensate is released from the combined trap, a matter-wave  
interference
pattern is
formed in few milliseconds.  The atom cloud released from one lattice
site expand, and overlap and interfere with atom clouds from neighboring
sites to form the robust interference pattern due to phase coherence. No
interference pattern can be formed without phase coherence. 
The pattern consists of a central peak at 
$y=0$ and two symmetrically spaced peaks, each containing about $10\%$ of 
total number of atoms,
moving apart from the central
peak \cite{cata,cata3b}.

The simulation of the formation of the  interference pattern
is performed
by loading the preformed condensate of figure  1 on two  lattices with
$\rho \le
25$ $\mu$m  and   
(a) $320$  $\mu$m  $\ge y \ge -320$  $\mu$m, and  (b)  $160$  $\mu$m $\ge
y \ge -160$
$\mu$m which
will permit the study of the
evolution of the interference pattern on a large interval  of space and
time. The interference pattern is formed   
by suddenly removing the combined
traps at time $t=0$.  The time evolution of the system 
is best illustrated via $P(y,t)$ and we plot in  figures 2 (a) and (b)
$P(y,t)$ vs. $y$ and $t$ for lattices  (a) and (b), respectively. 
The dynamics is 
symmetric about $y=0$ and only
$P(y,t)$ for positive $y$ is plotted in figure  2 (a). In these plots we
can
clearly see
the central condensate and the moving interference peak(s). The peaks
spread unobservably  slowly as they propagate, even 
after
reflection from
the boundary or after crossing each other. The phase coherence between 
the components of BEC at different sites of optical lattice is responsible
for the generation of the interference
pattern with very little or practically no spreading. 
 Without the
initial phase
coherence over a large number of lattice sites,
a repulsive condensate in the absence of a trap will disappear
in  few milliseconds \cite{ad}.  Each of the moving interference peaks
is  similar to atom laser \cite{1,al} which can be used in the scattering
of
two
coherent BECs and other purposes.

We have also examined the wave
function 
$\varphi(\hat \rho,\hat y,t)$ at different times (not reported
here). We find that
 there is virtually no spreading of the wave function in the axial $y$
direction during few hundred milliseconds.
 The phase coherence in the axial direction due to the optical
lattice is responsible for the localization of the peaks.

\begin{figure}
 
\begin{center}
\includegraphics[width=1.\linewidth]{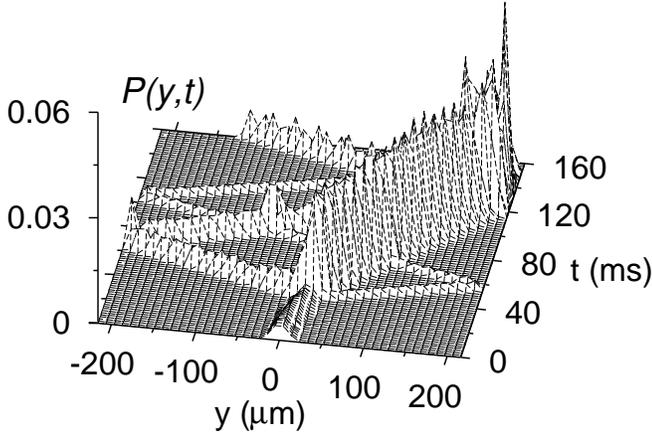}
\end{center}

\caption{$P(y,t)$
vs. $y$ and $t$ for an oscillating BEC on optical lattice after a
displacement of the magnetic trap through 25 $\mu$m along the optical axis
and 
upon the
removal of the combined traps at $t=35$ ms (hold time).}
 
\end{figure}

\begin{figure}
 
\begin{center}
\includegraphics[width=1.\linewidth]{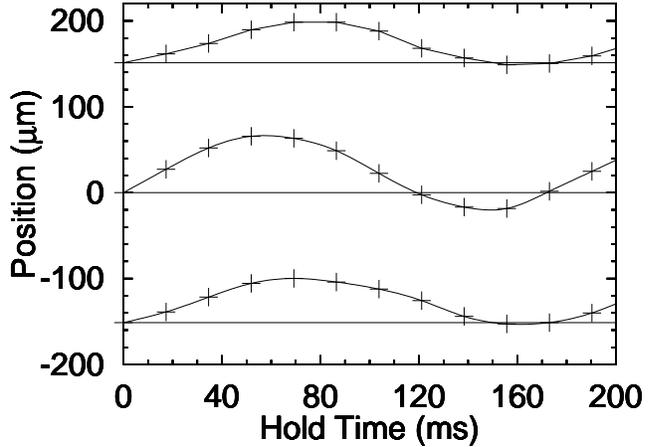}
\end{center}

\caption{Center of mass positions of the three interference peaks of the
expanded condensate after 20 ms of  free expansion vs. hold time of
the oscillating BEC. The magnetic trap is displaced through 
a distance of
25 $\mu$m. 
The $+$ symbols denote the results of simulation
which are joined by full lines to show the correlated oscillation of
the three peaks. }

\end{figure}

Next we consider an oscillating BEC in the combined harmonic 
and  optical traps. If we suddenly displace the magnetic trap along the
lattice axis by a small distance after the
formation of the BEC in the
combined trap, the condensate will be out of equilibrium and start to
oscillate. As the height of the potential-well barriers on the optical
lattice  
is much larger than the energy of the system, the atoms in the condensate
will move  by tunneling through the potential barriers.
This fluctuating transfer of Rb atoms across the potential barriers
is due to Josephson effect in  a neutral quantum liquid.
The experiment of Cataliotti {\it et al.} \cite{cata,cata3b} demonstrates
that 
the phase coherence between
different wells of the condensate is maintained during this mass transfer 
and  a matter-wave interference pattern with three peaks
is formed after the
removal
of
the joint trap. 
The peaks of the expanded
condensate oscillate in phase, thus showing that the quantum mechanical
phase coherence is maintained over the entire condensate.  
They studied
this problem experimentally in some detail by
varying  the time of
oscillation  of the BEC (hold time) before removing the combined trap
\cite{cata}.

To see if the present simulation can represent the essential features of
the phase coherence of the oscillating BEC, 
we load the GP equation with the BEC of the combined
harmonic oscillator and optical traps on a lattice defined by $200$ $\mu$m 
$\ge y
\ge -200$ $\mu$m  and $\rho \le 25$ $\mu$m and suddenly displace the
harmonic
trap along the
optical axis by  25 $\mu$m. The BEC
starts to oscillate and we allow the oscillation to evolve through a
certain interval of time, called hold time, before the removal of the
combined traps. The interference pattern is observed after some time of
free expansion and the positions of the interference peaks are noted. In
figure  3 we plot the one-dimensional probability $P(y,t) $ vs. $y$ and
$t$
after an initial evolution of the oscillation during 35
ms and observe the interference pattern for 160 ms. In
this case, unlike in  figures 2, the large central peak does not stay at
rest
and the sizes and positions of the two smaller peaks are not symmetrical
around $y=0$.  The clear formation of the interference pattern with very
little spreading even after reflection at the boundaries
and its
propagation for more than 160 ms is noted in figure  3 which
confirms the phase coherence.

\begin{figure}
 
\begin{center}
\includegraphics[width=1.\linewidth]{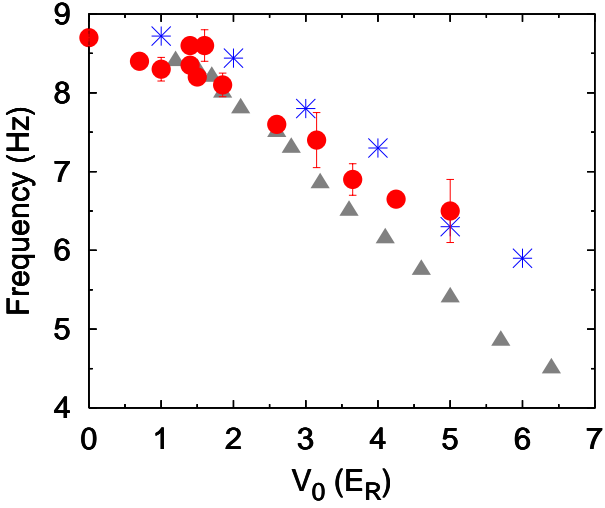}
\end{center}

\caption{The frequency of the atomic current in the array of Josephson
junctions as a function of optical lattice strength: $\bullet$
 with
error bar $-$ experiment of Cataliotti {\it et al.} \cite{cata}; 
$\triangle$    $-$ tight binding calculation \cite{cata}; $\star$
$-$ present
calculation.}

\end{figure}

\begin{figure}
 
\begin{center}
\includegraphics[width=1.\linewidth]{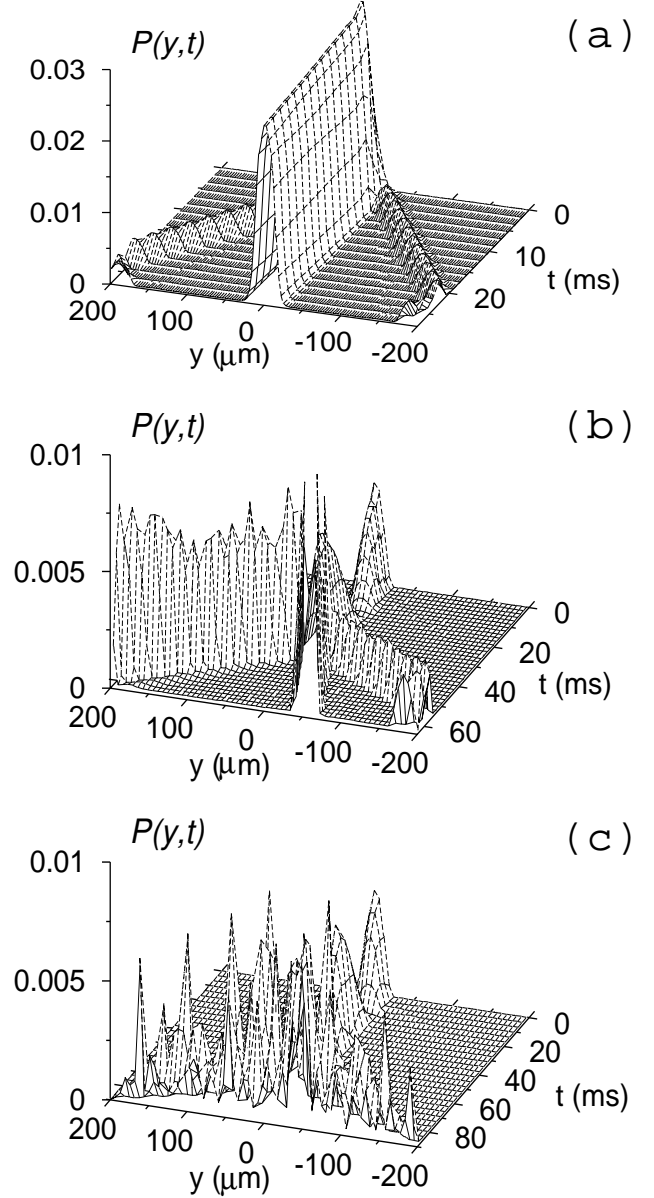}

\end{center}

\caption{$P(y,t)$
vs. $y$ and $t$ for a BEC on optical lattice after a sudden
displacement of the magnetic trap through 120 $\mu$m along the optical
axis
and 
upon the
removal of the combined traps after hold times (a) 0, (b) 35  ms, and
(c) 70  ms. The time evolution  is stopped upon 27.8 ms of free
expansion after the removal of the combined traps. }

\end{figure}

To study the phase coherence in detail we plot in figure  4 the positions
of
the expanded interference peaks after 20 ms
of free expansion
for  different hold times of oscillation in the displaced harmonic
potential. We find that the
interference peaks oscillate in phase showing the phase coherence in the
oscillating BEC. Similar oscillation was also 
observed in the experiment of Cataliotti {\it et al.} \cite{cata}.
From figure  4 we find that the period of this oscillation is about 
170
ms corresponding to a frequency of 5.9 Hz, which is 
very close to the experimental result exhibited in figure  3 of reference 
  \cite{cata}. To make a more complete comparison with  figure  3 of
reference 
  \cite{cata} we calculated the frequency of atomic current in the
array of Josephson junctions 
for different $V_0  $ and the
results are shown in figure  5 where we plot the present frequencies as
well
as those of the experiment  of Cataliotti {\it et al.}  \cite{cata} and of
their
tight-binding calculation.  From  figure 5 we see that the complete
solution of the GP equation has led to results in 
agreement with the 
experiment of
Cataliotti {\it et al.}. 
The agreement of the present calculation in  figure 5 performed with a
smaller condensate with experiment demonstrates that the frequency of
atomic current is mostly determined by the strength of the optical lattice
strength and is reasonably independent of the size of the condensate.

Finally, we consider the destruction of superfluidity in the condensate 
when the center of the magnetic trap is displaced 
along the optical lattice  by a distance larger than the critical
distance and
the BEC is allowed to stay in this displaced trap for an interval of time
(hold time). In this case the BEC does not execute an oscillatory motion
but its center moves very slowly  towards the new  center of the magnetic
trap.  
The destruction of  superfluidity and phase coherence for a larger hold
time in the
displaced trap manifests 
in the disappearance of the interference pattern upon free expansion as
noted in experiment \cite{cata2}. As in that experiment, we consider a
displacement of the magnetic trap through 120 $\mu$m and allow the
condensate to freely expand for 27.8 ms after different hold times in the 
optical and displaced magnetic traps.

For numerical simulation we load the BEC of  figure 1 on a lattice with 
$\rho \le 25$ $\mu$m and 200 $\mu$m $\ge y\ge$ $-200$ $\mu$m and study the
its evolution  after an initial displacement of  120 $\mu$m of
the magnetic trap for hold times 0, 35 ms, and 70 ms. As in the experiment
no oscillatory motion of the BEC is noted in the displaced trap. 
The corresponding
probability densities are plotted in  figures 6 (a), (b), and (c),
respectively. For hold times 0 and 35 ms prominent interference pattern is
formed upon free expansion as we can see in  figures  6 (a) and (b). 
In these cases three separate pieces of interference patterns  
corresponding  to three distinct trails can be identified. 
However,
as the hold time in the displaced trap increases the maxima
of the interference pattern
mixes up and finally for a hold time of 
70 ms the interference pattern is
completely destroyed as we find in  figure 6 (c) in agreement with the
experiment \cite{cata2}. 

As the BEC is allowed to evolve for a substantial  interval of time after
a large displacement of the magnetic  trap along the optical axis a
dynamical instability of classical nature sets in and the system can not
evolve maintaining the phase coherence \cite{sm,cata2}. This has been
explicitly demonstrated in the present simulation which results in the
destruction of the interference pattern.

\section{Conclusion}

In conclusion, to understand theoretically the experiments by Cataliotti
{\it et al.} \cite{cata,cata2}, we have studied in detail the phase
coherence
along a cigar-shaped condensate loaded in a combined axially-symmetric
harmonic trap and  optical lattice trap using the
solution of the mean-field  GP equation.  Upon removal of the combined
traps, the
formation of an  interference pattern clearly demonstrates the
phase coherence over a very large number of optical lattice sites. 
Each of the moving interference peaks formed of coherent matter wave
is similar to a atom laser observed experimentally
\cite{1,al}.
 The phase coherence along the optical lattice axis of the condensate
is maintained even if the initial BEC
is set into oscillation by suddenly shifting the harmonic trap along the
optical axis through a small distance and keeping the BEC in the
displaced trap for a certain hold time.  This is clearly demonstrated by
the noted correlated oscillation
of the condensate peaks after free expansion for different hold
times.  The present mean-field model provides a proper account of the
frequency of atomic current in the array of Josephson Junctions 
in  agreement  
with experiment  \cite{cata}. 

However, if the initial
displacement of the harmonic trap along the optical axis is larger than a
critical value and the BEC is maintained in the displaced trap for a
certain time, the phase coherence is destroyed. Consequently,
after release from the combined trap no interference pattern is formed in
agreement with experiment \cite{cata2}.

\vskip 1cm

\noindent{The work was supported in part by the CNPq and FAPESP
of Brazil.}


 \end{document}